\documentclass{iau}
\usepackage{graphicx}

\title[Magnetic field of $\eta$ Aql] 
{Magnetic field of the classical Cepheid $\eta$ Aql: new results}

\author[Butkovskaya et al.]   
{V.Butkovskaya$^1$, S.Plachinda$^1$, D.Baklanova$^1$
 \and V.Butkovskyi$^2$}

\affiliation{$^1$Crimean Astrophysical Observatory of Taras Shevchenko National University of Kyiv, \\ 98409,  Nauchny, Crimea, Ukraine,\\ email: {\tt varya@crao.crimea.ua} \\[\affilskip]
$^2$Taurida National V.I.Vernadsky University, \\ 95007, Vernadskogo str. 4, Simferopol, Crimea, Ukraine}

\pubyear{2013}
\volume{302}  
\pagerange{119--126}
\jname{Magnetic fields throughout stellar evolution}
\editors{M.Jardine, P.Petit, \& H.Spruit, eds.}
\begin{document}

\maketitle

\begin{abstract}
We present the results of the spectropolarimetric study of the classical Cepheid $\eta$ Aql in 2002, 2004, 2010, and 2012. The longitudinal magnetic field of $\eta$ Aql was found to be variable with the pulsation cycle of 7.176726 day. The amplitude, phase, and mean value of the field vary from year to year presumably due to stellar rotation or dynamo mechanisms. 
\end{abstract}

\firstsection 
\section{Introduction}

Historically the first photoelectric magnetometer observations of $\eta$ Aql have been performed by 
\cite[Borra et al. (1981)]{borra81},
and 
\cite[Borra et al. (1984)]{borra84}. 
These studies detected no magnetic field on the star with the best error 7.7 G.  The presence of a magnetic field on $\eta$ Aql was firstly reported by 
\cite[Plachinda (2000)]{[plach00},
who found that in 1991 the longitudinal component of the magnetic field was variable from -100 to +50 G during pulsation cycle. 
\cite[Wade et al. (2002)]{wade02}
detected no convincing evidence of a photospheric magnetic field on $\eta$ Aql during three nights in 2001. They phased obtained magnetic field values with the pulsation period and noted that around phase 0.8 their results are strongly inconsistent with those of 
\cite[Plachinda (2000)]{[plach00}. 
The authors concluded that $\eta$ Aql is a non-magnetic star, at least at the level of 10 G. More recently 
\cite[Grunhut et al. (2010)]{grunhut04}
detected clear Zeeman signatures in Stokes V for 9 later-type supergiants (including $\eta$ Aql). Firstly the presence of the magnetic field on later-type supergiants was strongly proved by 
\cite[Plachinda (2005)]{plach05},
who detected the longitudinal magnetic field on two yellow supergiants - $\epsilon$ Gem and $\epsilon$ Peg. So while the presence of a magnetic field on later-type supergiants (including pulsating classical Cepheids) can be considered proven, a doubt on a variability of the magnetic field with the stellar pulsations is shared yet by many astronomers. We present here the new results of the spectropolarimetric observations of $\eta$ Aql.

\section{Observations}

Spectropolarimetric observations of $\eta$ Aql have been performed in the spectral region 6210 - 6270 \AA ~during 60 nights from 2002 to 2012 using coude spectrograph at the 2.6-m Shajn telescope mounted at the Crimean Astrophysical Observatory (CrAO, Ukraine). The longitudinal magnetic field was calculated using the technique which is described in detail by 
\cite[Butkovskaya \& Plachinda (2007)]{butk07}. Our technique of the spectropolarimetric measurements and magnetic field calculation allow us to exclude spurious magnetic signal produced by variation of spectral line profiles due to a pulsation. 

\section{Results}
\begin{figure}[b]
 \vspace*{-1.0 cm}
\begin{center}
 \includegraphics[width=4.2in]{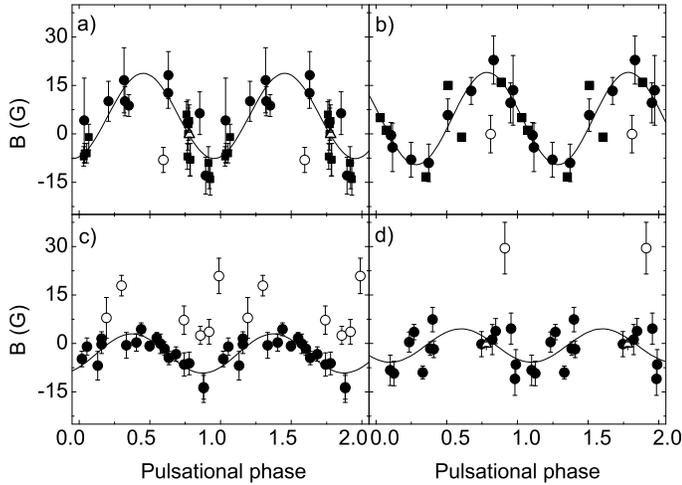} 
 \caption{Longitudinal magnetic field of $\eta$ Aql folded in phase with the 7.176726 day pulsation period: \textbf{a)} \textit{closed and open circles}: our data obtained in 2002, \textit{black squares}: data by 
\cite[Wade et al. (2002)]{wade02}, 
\textit{open triangles}: data by 
\cite[Grunhut et al. (2010)]{grunhut04}; 
\textbf{b)} \textit{closed and open circles}: our data obtained in 2004, \textit{black squares}: data by 
\cite[Borra et al. (1981)]{borra81}, 
and
\cite[Borra et al. (1984)]{borra84},
(mean $\sigma$ = 13 G); \textbf{c)} \textit{closed and open circles}: our data obtained in 2010; \textbf{d)} \textit{closed and open circles}: our data obtained in 2012, and \textit{open triangles}: data by 
\cite[Grunhut et al. (2010)]{grunhut04}. 
Fitting sinusoids are shown by strong lines. Open circles represent our data that have not been taken into account for the fits; we suppose that the data are deviated from the common curve because of an unknown yet reason, for example, shock waves or buoyantly ~rising magnetic ~flux ~tubes.}
   \label{fig1}
\end{center}
\end{figure}

The pulsation modulation of the longitudinal magnetic field of $\eta$ Aql in different years is illustrated in Fig.\,\ref{fig1}. Our data are supplemented with data of 
\cite[Borra et al. (1981)]{borra81}, 
\cite[Borra et al. (1984)]{borra84},
\cite[Wade et al. (2002)]{wade02},
\cite[Grunhut et al. (2010)]{grunhut04}
All data are folded in phase according to the pulsation ephemeris $JD = 2450100.861+7.176726E$ by 
\cite[Kiss \& Vinko (2000)]{kiss00}. 

The pulsation modulation of the longitudinal magnetic field of $\eta$ Aql is not unique: 
\cite[Butkovskaya \& Plachinda (2007)]{butk07}
discovered variability of the magnetic field of $\gamma$ Peg (B2 IV) with the radial pulsation period and described possible origins of this variability.

\end{document}